\RequirePackage{lineno} 
\documentclass[ALICE,manyauthors,preprintnumbers]{cernphprep}

\def\pt{\mbox{$p_{\rm T}$ }}   
\def\snn{\mbox{$\sqrt{s_{\rm NN}}$}}   
\def\pb {Pb-Pb }
\def\au {Au-Au }

\def\j {\mbox{J/$\psi$ }}
\def\vv {$v_2$ }
\def\vpt{\mbox{$v_2(p_{\rm T})$ }}
\def\raa {\mbox{$R_{\rm{AA}}$}}

\newcommand{ \be }{\begin{eqnarray}}
\newcommand{ \ee }{\end{eqnarray}}

\usepackage[sort&compress,numbers]{natbib}
\usepackage{doi}
\usepackage{hyperref}
\usepackage{lineno}

\begin{document}%
%%%%%%%%%%%%% ptdr definitions %%%%%%%%%%%%%%%%%%%%%
%

%%%%%%%%%%%%%%%  Title page %%%%%%%%%%%%%%%%%%%%%%%%
%
\begin{titlepage}
\PHnumber{2013-042}                 % required, obtained from PH
\PHdate{March 20, 2013}             % required
%\EXPnumber{ALICE-INT-2010-9999}     % optional
%\EXPdate{12 October 2010}           % optional
%
%
%%% Put your own title + short title here:
\title{\j Elliptic Flow in \pb Collisions at \snn~ = 2.76 TeV}
\ShortTitle{\j Elliptic Flow in \pb Collisions at \snn~ = 2.76 TeV}   % appears on right page headers
%
%%% Do not change the next lines!
\Collaboration{ALICE Collaboration%
         \thanks{See Appendix~\ref{app:collab} for the list of collaboration
                      members}}
\ShortAuthor{ALICE Collaboration}      % appears on left page headers, do not change
\begin{abstract}
%!TEX root = Jpsiv2_PRL.tex
 We report on the first measurement of inclusive \j elliptic flow, $v_{2}$, in heavy-ion collisions at the LHC.  
The measurement is performed with the ALICE detector in \pb collisions at \snn~=~2.76 TeV in the rapidity range $2.5<y<4.0$. 
The dependence of the \j \vv on the collision centrality and on the \j transverse momentum is studied in the range  $0 \leq p_{\rm T} < 10$~GeV/$c$.
For semi-central \pb collisions at \snn~=~2.76 TeV, an indication of non-zero \vv is observed with a maximum value of $v_{2}=0.116\pm0.046(\rm{stat.})\pm0.029(\rm{syst.})$ for \j in the transverse momentum range $2 \leq p_{\rm T} < 4$~GeV/$c$.
The elliptic flow measurement complements the previously reported ALICE results on the inclusive \j nuclear modification factor and favors the scenario of a significant fraction of J/$\psi$ production from charm quarks in a deconfined partonic phase.  
\end{abstract}
\end{titlepage}
%\newpage
\setcounter{page}{2}
%
%\modulolinenumbers[5]
%\linenumbers
%!TEX root = Jpsiv2_CernPreprint.tex
Ultra-relativistic heavy-ion collisions enable the study of matter at high temperature and pressure where Quantum Chromodynamics predicts the existence of a deconfined state of partonic matter, the Quark-Gluon Plasma (QGP). 
Heavy quarks are expected to be produced in the primary partonic scatterings and to interact with this partonic medium making them ideal probes of the QGP. 
Quarkonia (a heavy quark and anti-quark bound state) are therefore expected to be sensitive to the properties of the strongly-interacting system formed in the early stages of heavy-ion collisions~\cite{Bedjidian:2004gd}. 
According to the color-screening model~\cite{Matsui:1986dk}, quarkonium states are suppressed in the medium with different dissociation probabilities for the various states. 
Recently, the CMS Collaboration at the Large Hadron Collider (LHC) reported about the observation of the sequential suppression in the $\Upsilon$ sector~\cite{Chatrchyan:2012lxa}.  
The ALICE Collaboration published the inclusive~\footnote{Inclusive J/$\psi$ include prompt J/$\psi$ (direct and decays from higher mass charmonium states) and non-prompt J/$\psi$ (feed down from b-hadron decays).} J/$\psi$ nuclear modification factor \raa~down to zero transverse momentum ($p_{\rm T}$) at forward rapidity in Pb-Pb collisions at \snn~=~2.76~TeV~\cite{Abelev:2012rv}. 
The \raa~compares the yields in \pb to those in pp collisions scaled by the number of binary nucleon-nucleon collisions. 
The inclusive \j \raa~reported is larger than that measured at the SPS \cite{Alessandro:2004ap} and at RHIC \cite {Adare:2006ns, Adare:2011yf} for central collisions and does not exhibit a significant centrality dependence. 
Complementarily, the CMS Collaboration measured the high $p_{\rm T}$ ($6.5 \leq p_{\rm T}<30$ GeV/$c$) prompt \j \raa~in the rapidity range $|y|<2.4$~\cite{Chatrchyan:2012np}.  
The CMS data show that high \pt \j are more suppressed than low \pt \mbox{J/$\psi$} and that this suppression does exhibit a strong centrality dependence.  

The low \pt \j \raa~can be qualitatively understood with models including full~\cite{BraunMunzinger:2000px,Andronic:2011yq} or partial~\cite{Zhao:2011cv, Liu:2009nb} regeneration of \j from deconfined charm quarks in the medium. 
This mechanism was first proposed by the statistical hadronization model, which assumes deconfinement and thermal equilibrium of the bulk of $c\bar{c}$ pairs to produce \j at the phase boundary by statistical hadronization only~\cite{BraunMunzinger:2000px}. 
Later, the transport models proposed a dynamical competition between the \j suppression by the QGP and the regeneration mechanism, which enables them to also describe the \j \raa~versus \pt \cite{Zhao:2011cv, Liu:2009nb}.  
More differential studies, like the \j elliptic flow, could help to assess the assumption of charm quark thermalization in the medium. 

The azimuthal distribution of particles in the transverse plane is also sensitive to the dynamics of the early stages of heavy-ion collisions.
In non-central collisions, the geometrical overlap region and, therefore, the initial matter distribution are anisotropic (almond-shaped).  
If the matter is strongly interacting, this spatial asymmetry is converted via multiple collisions into an anisotropic momentum distribution~\cite{Ollitrault:1992bk}.  
The second coefficient of the Fourier expansion describing the final state particle azimuthal distribution with respect to the reaction plane, $v_2$, is called elliptic flow. 
The reaction plane is defined by the beam axis and the impact parameter vector of the colliding nuclei.

Within the transport model scenario~\cite{Zhao:2011cv, Liu:2009nb} observed \j have two origins. 
First, primordial J/$\psi$ produced in the initial hard scatterings, traverse and interact with the created medium. During this process they may be dissociated.
Second, \j could be regenerated from deconfined charm quarks in the QGP. 
Primordial \j emitted in-plane traverse a shorter path through the medium than those emitted out-of-plane resulting in a small azimuthal anisotropy for the surviving \mbox{J/$\psi$}. 
Regenerated \j inherit the elliptic flow of the charm quarks in the QGP. 
If charm quarks do thermalize in the QGP, then \j formed there can exhibit a large elliptic flow. In the calculation by Zhao {\it et al.}~\cite{Zhao2013611c}, the \vv of \j at \pt $\approx$ 2.5 GeV/c is 0.02 and 0.2 for primordial and regenerated \j, respectively.

At RHIC, the (preliminary) measurements by the (PHENIX) STAR Collaboration of the \j \vv in \au collisions at \snn~=~200~GeV~\cite{Silvestre:2008tw,Adamczyk:2012pw} are consistent with zero albeit with large uncertainties in the \pt and centrality ranges  (0--5 GeV/$c$) 2--10 GeV/$c$ and ($20\%$--$60\%$) $10\%$--$40\%$. 
In Pb-Pb collisions at the LHC, the higher energy density of the medium should favor the charm quark thermalization, and thus increase its flow. In addition, the large number of $c\bar{c}$ pairs produced should favor the formation of \j by regeneration mechanisms. Both effects should lead to an increase of the \vv of the observed \j.
%The quarkonium \vv measurement is especially promising at the LHC where the high energy density of the medium and the large number of  $c\bar{c}$ pairs produced in Pb-Pb collisions should favor the development of flow and the regeneration mechanisms.

In this Letter, we report ALICE results on inclusive \j elliptic flow in \pb collisions at \snn~=~2.76 TeV at forward rapidity,  measured via the $\mu^{+} \mu^{-}$ decay channel. The results are presented as a function of transverse momentum and collision centrality.

The ALICE detector is described in~\cite{Aamodt:2008zz}. 
At forward rapidity  ($2.5 < y < 4$) the production of quarkonia is measured in the muon spectrometer~\footnote{In the ALICE reference frame, the muon spectrometer covers a negative $\eta$ range and consequently a negative $y$ range.
We have chosen to present our results with a positive $y$ notation.} down  to $p_{\rm{T}} = 0$. 
The spectrometer consists of an absorber stopping the hadrons in front of five tracking stations comprising two planes of cathode pad chambers each, with the third station inside a dipole magnet. 
The tracking apparatus is completed by a triggering system made of four planes of resistive plate chambers downstream of an iron wall,  which absorbs secondary hadrons escaping from the front absorber and low momentum muons.
Also used in this analysis are two cylindrical layers of silicon pixel detectors, to determine the location of the interaction point, and two scintillator arrays (VZERO). 
The VZERO counters consist of two arrays of 32 scintillator sectors each distributed in four rings covering  $2.8 \leq \eta \leq 5.1$ (VZERO-A) and $-3.7 \leq \eta \leq -1.7$ (VZERO-C).
All of these detectors have full azimuthal coverage.
The data sample used for this analysis, collected in 2011, amounts to $17 \times 10^{6}$ dimuon unlike sign (MU) triggered Pb-Pb collisions and corresponds to an integrated luminosity $\mathcal{L}_{\rm int} \approx 70$  $\mu$b$^{-1}$. 
The MU trigger requires a minimum bias (MB) trigger and at least a pair of opposite-sign (OS) track segments, each with a \pt above the threshold of the on-line trigger algorithm. 
This \pt threshold was set to provide 50\% efficiency for muon tracks with \pt = 1 GeV/$c$. 
The MB trigger requires a signal in both VZERO-A and VZERO-C.
The beam-induced  background was further reduced offline using the VZERO and the zero degree calorimeter (ZDC) timing information. 
The contribution from electromagnetic processes was removed by requiring a minimum energy deposited in the neutron ZDCs~\cite{ALICE:2012aa}.
The centrality determination is based on a fit of the VZERO amplitude distribution~\cite{PhysRevLett.106.032301,Abelev:2013qoq}.
The average number of participating nucleons $\langle N_{\rm{part}} \rangle$ for the centrality classes used in this analysis (see Table.~\ref{tab:res}) are derived from a Glauber model calculation~\cite{PhysRevLett.106.032301,Abelev:2013qoq}. 

\begin{figure}
\includegraphics[width=0.9\linewidth]{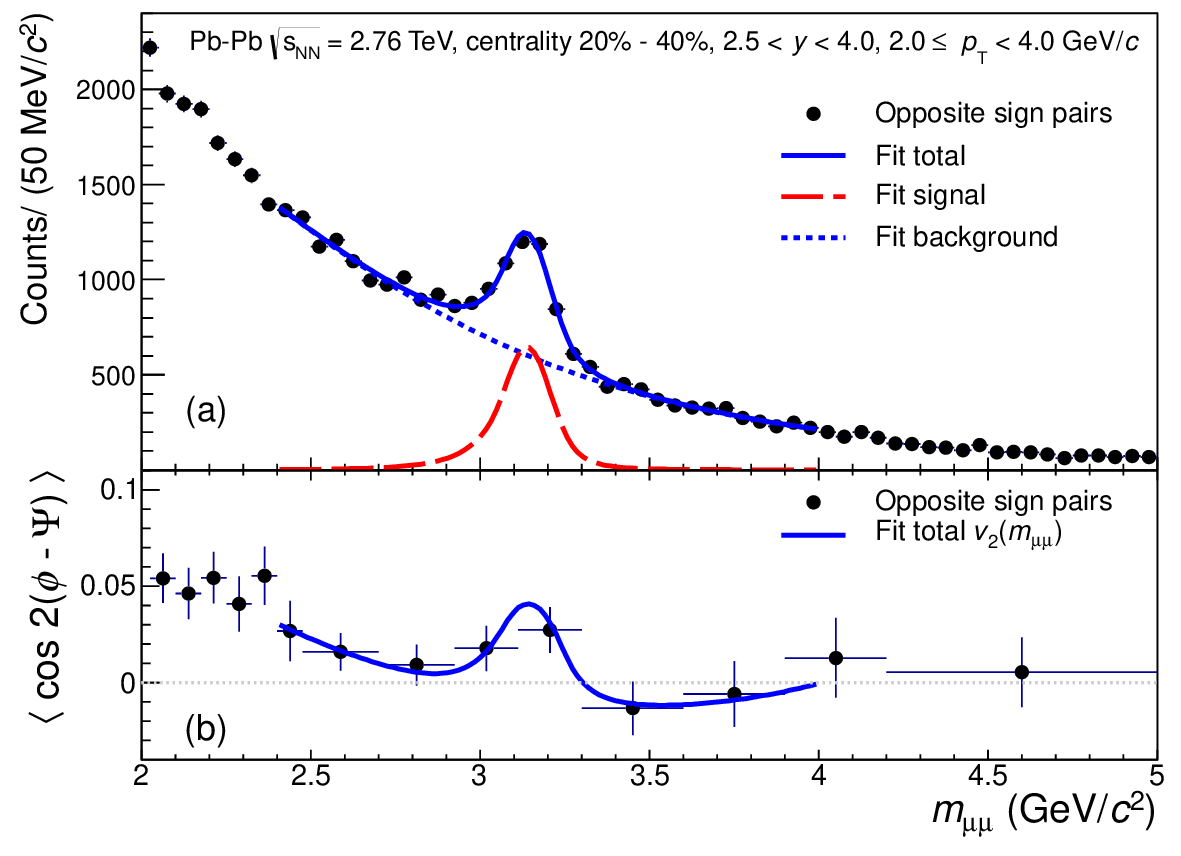} %PRL
\caption{\label{fig:massv2} (color online) Invariant mass distribution (a)  and $\langle \cos 2 (\phi - \Psi) \rangle$ as a function of $m_{\mu\mu}$ (b) of OS dimuons with $2 \leq p_{\rm T}<4$ GeV/$c$  and $2.5 < y < 4$ in semi-central (20\%--40\%) Pb-Pb collisions.}
\end{figure}

J/$\psi$ candidates are formed by combining pairs of OS tracks reconstructed in the geometrical acceptance of the muon spectrometer. 
To improve the muon identification, the reconstructed tracks in the tracking chambers are required to match a track segment in the trigger system above the \pt threshold aforementioned.

The \j \vv is calculated using event plane (EP) based methods.  
The azimuthal angle $\Psi$ of the second harmonic EP is used to estimate the reaction plane angle~\cite{Poskanzer:1998yz}.  
$\Psi$ is determined from the azimuthal distribution of the VZERO amplitude. 
A two step flattening procedure of the EP azimuthal distribution was applied as described in~\cite{Selyuzhenkov:2007zi} and~\cite{Barrette:1997pt}, respectively. It results in an EP azimuthal distribution uniform to better than 2$\%$ for all centrality classes under study. 
The VZERO-C has a common acceptance region with the muon spectrometer. 
Therefore, only the VZERO-A was used for the EP determination to avoid autocorrelations.
The \j $v_{2}$ results were obtained determining $v_{2} = \langle \cos 2 (\phi - \Psi) \rangle$ versus the invariant mass ($m_{\mu\mu}$)~\cite{Borghini:2004ra}, where $\phi$ is the OS dimuon azimuthal angle. 
The resulting $v_{2}(m_{\mu\mu})$ distribution is fitted using
\begin{equation}\label{eq:v2Tot}
v_{2}(m_{\mu\mu}) = v_{2}^{\rm sig}\alpha(m_{\mu\mu}) + v_{2}^{\rm bkg}(m_{\mu\mu})[1-\alpha(m_{\mu\mu})],
\end{equation}
where $v_{2}^{\rm sig}$ and $v_{2}^{\rm bkg}$ correspond to the $v_{2}$ of the \j signal and of the background, respectively (see Fig.~\ref{fig:massv2} (b)). $v_{2}^{\rm bkg}$ was parametrized using a second order polynomial. 
Here, $\alpha(m_{\mu\mu}) = S/(S+B)$ is the ratio of the signal over the sum of the signal plus background of the $m_{\mu\mu}$ distributions. 
It is extracted from fits to the OS invariant mass distribution (see Fig.~\ref{fig:massv2} (a)) in each \pt and centrality class. 
The J/$\psi$ line shape was described with a Crystal Ball (CB) function and the underlying continuum with either a third order polynomial or a Gaussian with a width linearly varying with mass.
The CB function connects a Gaussian core with a power-law tail~\cite{CBdef} at low mass to account for energy loss fluctuations and radiative decays. 
An extended CB function with an additional power-law tail at high mass, to account for alignment and calibration biases, was also used.
The combination of several CB and underlying continuum parametrizations described before were tested to assess the signal and the related systematic uncertainties. 
The \j \vv and its statistical uncertainty in each \pt and centrality class were determined as the average of the $v_{2}^{\rm sig}$ obtained by fitting $v_{2}(m_{\mu\mu})$ using Eq.~\ref{eq:v2Tot} with the various $\alpha(m_{\mu\mu})$, while the corresponding systematic uncertainties were defined as the RMS of these results.
Figure~\ref{fig:massv2} shows typical fits of the OS invariant mass distribution (a) and of the $\langle \cos 2 (\phi - \Psi) \rangle$ as a function of $m_{\mu\mu}$ (b) in the 20\%--40\% centrality class.
The procedure above was repeated using either a first order polynomial or its inverse as $v_{2}^{\rm bkg}$ parametrization. 
The largest deviation of the results obtained with the three different $v_{2}^{\rm bkg}$ parametrizations was conservatively adopted as the systematic uncertainty related to the unknown shape of the $v_{2}^{\rm bkg}(m_{\mu\mu})$.
This turns out to be often the dominant source of systematic uncertainties with the uncertainty from the signal extraction being the second one. 
It was checked that different choices of invariant mass binnings yield \vv values that are consistent within uncertainties.
A similar method was used to extract the uncorrected (for detector acceptance and efficiency) average transverse momentum ($\langle p_{\rm T} \rangle^{\rm uncor}$) of the reconstructed J/$\psi$ in each centrality and \pt class. 
The $\langle p_{\rm T} \rangle^{\rm uncor}$ is used to locate the data points when plotted as a function of $p_{\rm T}$. 
Consistent \vv values were obtained using an alternative method~\cite{Poskanzer:1998yz} in which the \j raw yield is extracted, as described before, in bins of $(\phi-\Psi)$ and $v_{2}$ is evaluated by fitting the data with the function $\frac{dN}{d(\phi-\Psi)} = A[1 + 2v_{2}\cos 2(\phi-\Psi)]$, where $A$ is a normalization constant. 
As an additional check the first analysis procedure~\cite{Borghini:2004ra} was also applied to the same-sign (SS) dimuons. 
As expected, no J/$\psi$ signal is seen in either the invariant mass distribution or the $\langle \cos 2 (\phi - \Psi) \rangle$ as a function of $m_{\mu\mu}$ of SS dimuons. 
In both cases the SS dimuons exhibit the same trend as the continuum of the OS dimuons.

The finite resolution in the EP determination smears out the azimuthal distributions and lowers the value of the measured anisotropy~\cite{Poskanzer:1998yz}. 
The VZERO-A EP resolution as a function of the centrality was determined using MB events and the 3 sub-event method~\cite{Poskanzer:1998yz}. 
To estimate the systematic uncertainty from the EP determination two sets of 3 sub-events were used: first, VZERO-A, VZERO-C and the time projection chamber (TPC), with pseudo-rapidity gaps  $\Delta \eta_{\rm V0A-TPC}$=1.9, $\Delta \eta_{\rm V0A-V0C}$=4.5 and $\Delta \eta_{\rm TPC-V0C}$=0.8; 
second, VZERO-A, ring 0 of VZERO-C and VZERO-C-3rd ring, with pseudo-rapidity gaps $\Delta \eta_{\rm V0A-V0C0}$=6.0, $\Delta \eta_{\rm V0C0-V0C3}$=1.0 and $\Delta \eta_{\rm V0A-V0C3}$=4.5. 
The differences between the EP resolution for VZERO-A obtained from these two sets of sub-events are taken as systematic uncertainties. 
Since \vv is measured here in a wide centrality class, the resolution must reflect the distribution of events with a \j within the class.
Therefore, the EP resolution for each wide class was calculated as the average of the values obtained in finer centrality classes weighted by the number of reconstructed $J/\psi$.  
Table~\ref{tab:res} shows the corresponding resolution for each centrality class which is applied to the results reported in this Letter.

%--------------------------------------------------
\begin{table}
\caption{\label{tab:res} $\langle N_{\rm{part}} \rangle$ and VZERO-A EP resolution for the centrality classes  expressed in percentages of the nuclear cross section~\cite{PhysRevLett.106.032301}. }
%\begin{ruledtabular}
\begin{tabular}{cccccc}
Centrality   & $\langle N_{\rm part} \rangle$  &  EP resolution $\pm$ (\rm{stat.}) $\pm$ (\rm{syst.})   \\
\hline
  5\%--20\%  &  283 $\pm$ 4 &   0.548 $\pm$ 0.003 $\pm$ 0.009 \\    
20\%--40\%  &  157 $\pm$ 3 &   0.610 $\pm$ 0.002 $\pm$ 0.008  \\
40\%--60\%  &    69 $\pm$ 2 &   0.451 $\pm$ 0.003 $\pm$ 0.008  \\
60\%--90\%  &    15 $\pm$ 1 &   0.185 $\pm$ 0.005 $\pm$ 0.013  \\
20\%--60\%  &  113 $\pm$ 3 &   0.576 $\pm$ 0.002 $\pm$ 0.008  \\
\end{tabular}
%\end{ruledtabular}
\end{table}
%--------------------------------------------------

The \j reconstruction efficiency depends on the detector occupancy, which could bias the \vv measurement. 
This effect was evaluated by embedding azimuthally isotropic simulated J/$\psi \rightarrow \mu^{+} \mu^{-}$ decays into real events. 
The measured \vv of those embedded \j does not deviate from zero by more than 0.015 in the centrality and \pt classes considered. 
This value is used as a conservative systematic uncertainty on all measured \vv values. 

\begin{figure}
\includegraphics[width=0.91\linewidth]{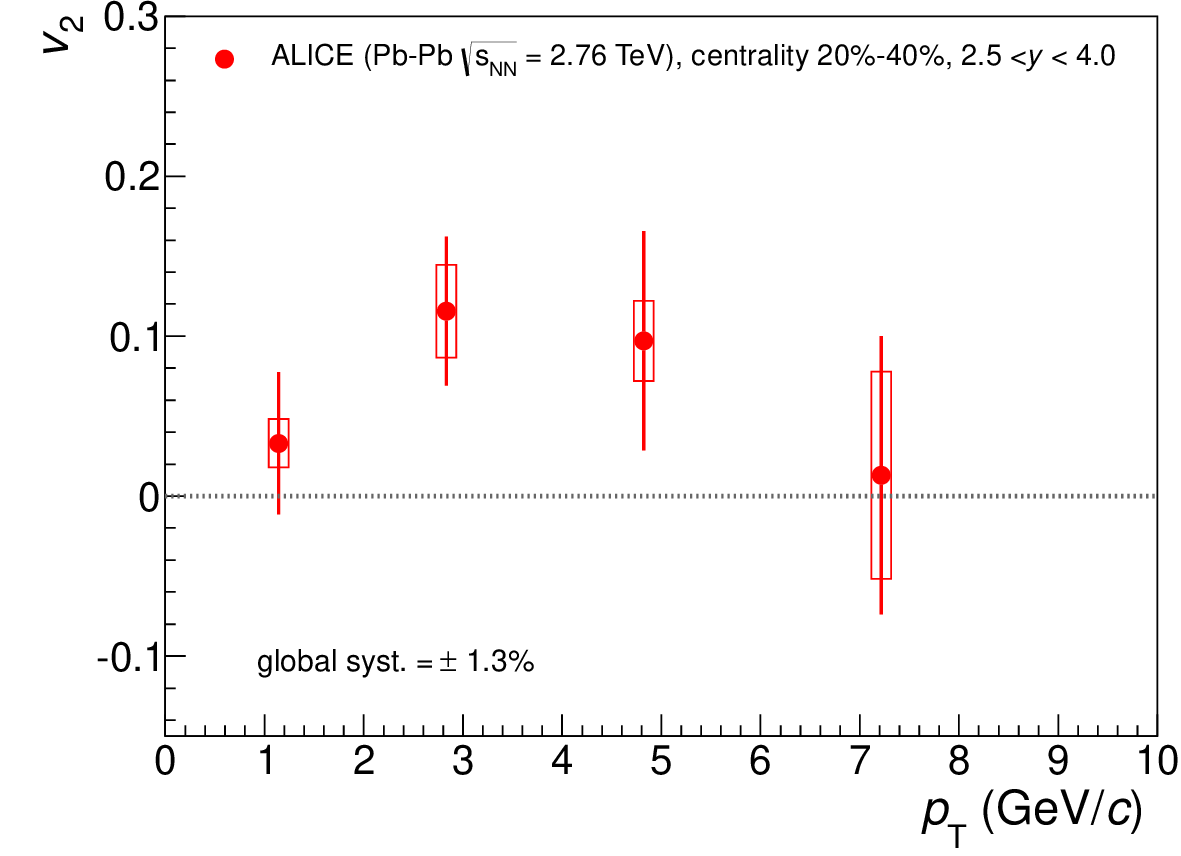}
\caption{\label{fig:v2pt} (color online) Inclusive \j \vpt  for semi-central (20\%--40\%) \pb collisions at \snn~=~2.76~TeV (see text for details on uncertainties). The used \pt ranges are: 0--2, 2--4, 4--6 and 6--10 GeV/$c$.}
\end{figure}

Figure~\ref{fig:v2pt} shows the \pt dependence of the inclusive \j \vv for semi-central (20\%--40\%) \pb collisions at \snn~=~2.76~TeV. 
The vertical bars show the statistical uncertainties while the boxes indicate the point-to-point uncorrelated systematic uncertainties, which include those from the signal extraction, the $v_2^{bkg}$ shape and the reconstruction efficiency.
The global correlated relative systematic uncertainty on the EP resolution is 1.3\%.  
A non-zero \vv is observed in the intermediate \pt range $2 \leq p_{\rm T}<6$~GeV/$c$. 
Including statistical and systematic uncertainties the combined significance of a non-zero \vv in this \pt range is $2.7 \sigma$. 
At lower and higher \pt the inclusive \j \vv is compatible with zero within uncertainties.

\begin{figure}
\includegraphics[width=0.95\linewidth]{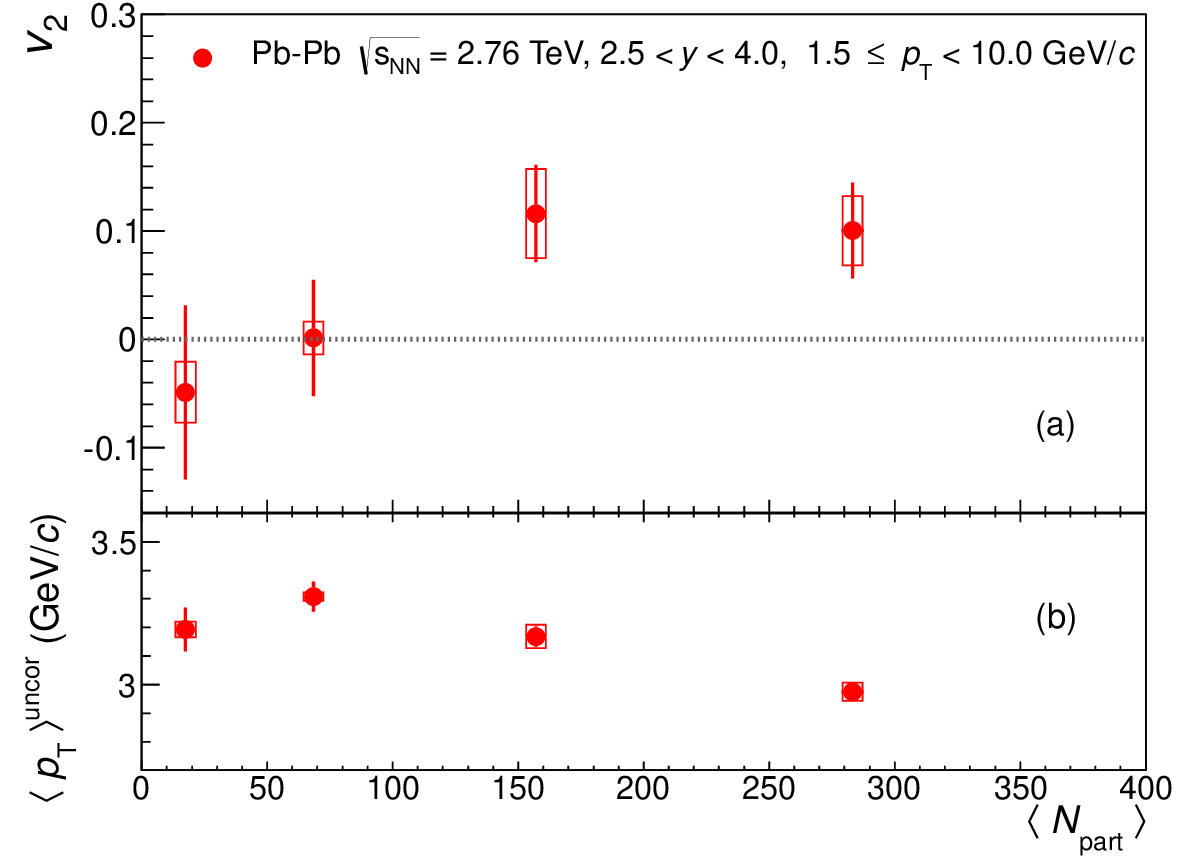}
\caption{\label{fig:v2cent2} (color online) \vv (a) and $\langle p_{\rm T} \rangle^{\rm uncor}$ (b) of inclusive \j with $1.5 \leq p_{\rm T} < 10$ GeV/$c$ as a function of $\langle N_{\rm{part}} \rangle$ in \pb collisions at \snn~=~2.76~TeV (see text for details on uncertainties).}
\end{figure}

%To study the centrality dependence of the \vv we select \j with $1.5 \leq p_{\rm T} < 10$ GeV/$c$ for which the signal to background ratio and the observed \vv are maximized. 
To study the centrality dependence of the \vv we select \j with $1.5 \leq p_{\rm T} < 10$ GeV/$c$. Indeed, below 1.5 GeV/$c$ the \vv of the \j is expected to be small \cite{Zhao2013611c} and the signal to background ratio is also low.
Since the initial spatial anisotropy for head-on collisions is small, the expected \vv is also small. 
In addition, for the 0\%--5\% centrality range the VZERO-A EP resolution is quite low and has higher systematic uncertainties.
Therefore, the 0\%--5\% centrality range was excluded.
Figure~\ref{fig:v2cent2} (a) shows \vv for inclusive \j with $1.5 \leq p_{\rm T} < 10$ GeV/$c$ as a function of $\langle N_{\rm{part}} \rangle$ in \pb collisions at \snn~=~2.76~TeV. 
Here, the point-to-point uncorrelated systematic uncertainties (boxes) also include, in addition to those discussed above, the uncertainty from the EP resolution determination. 
The measured \vv depends on the \pt distribution of the reconstructed \mbox{J/$\psi$}, which could vary with the collision centrality.
Therefore, $\langle p_{\rm T} \rangle^{\rm uncor}$ of the reconstructed \j is also shown in Fig.~\ref{fig:v2cent2} (b). 
The error bar indicates the statistical uncertainties while the boxes show the systematic uncertainties due to the \j signal extraction.
For the most central collisions, 5\%--20\% and 20\%--40\% the inclusive \j \vv for $1.5 \leq p_{\rm T} < 10$ GeV/$c$ are $0.101\pm0.044 (\rm{stat.})\pm0.032(\rm{syst.})$ and  $0.116\pm0.045 (\rm{stat.})\pm0.041(\rm{syst.})$, respectively. 
The combined significance of a non-zero \vv is 2.9$\sigma$. 
For more peripheral \pb collisions, the \vv is consistent with zero within uncertainties. 
Although there is a small variation with centrality, the $\langle p_{\rm T} \rangle^{\rm uncor}$ stays in the range 3.0--3.3~GeV/$c$, indicating that the bulk of the reconstructed \j are in the same \pt range for all centralities. 
Thus, the observed centrality dependence of the \vv for inclusive \j with $1.5 \leq p_{\rm T} < 10$~GeV/$c$ does not result from any bias in the sampled \pt distributions.
For \j with $p_{\rm T}<1.5$~GeV/$c$ (not shown), the \vv is compatible with zero within one standard deviation for the four centrality classes. 
The $\langle p_{\rm T} \rangle^{\rm uncor}$ ranges from about 0.75 to 0.9~GeV/$c$. 

\begin{figure}
\includegraphics[width=0.95\linewidth]{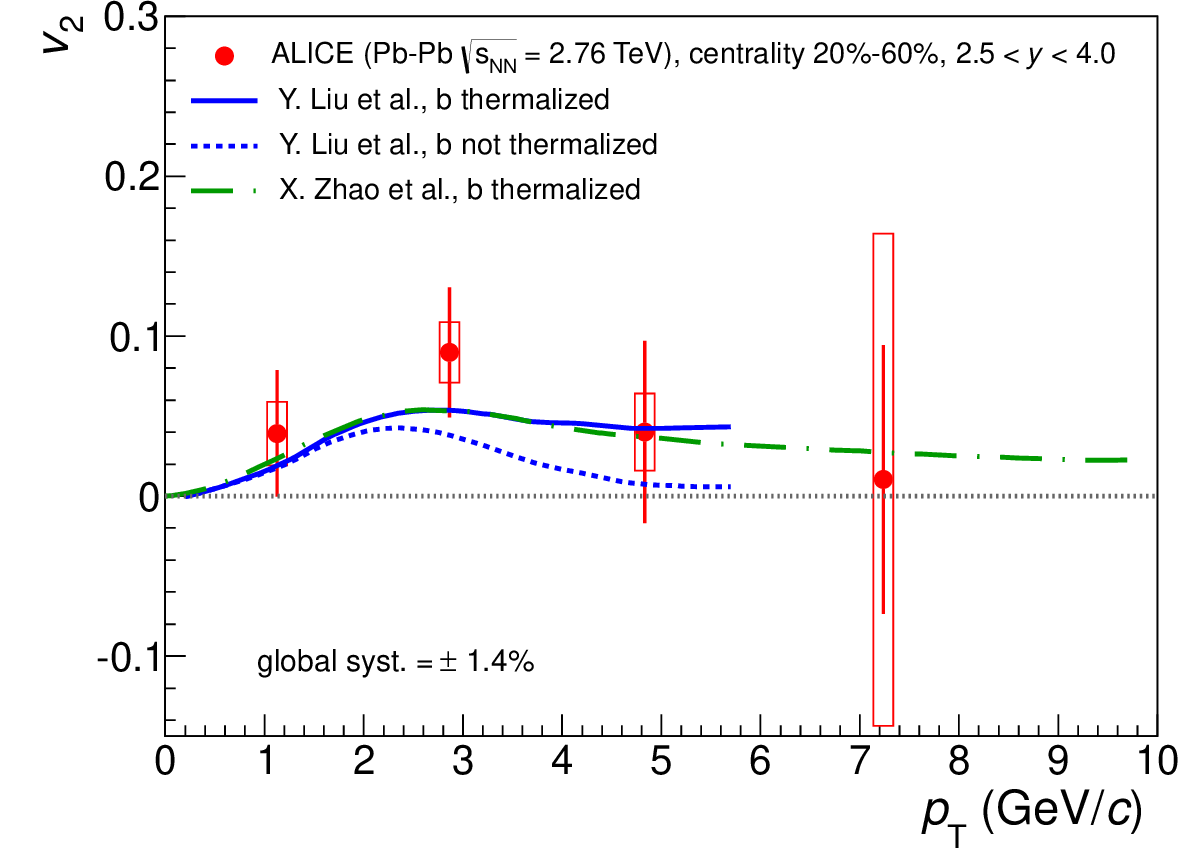}
\caption{\label{fig:v2ptcomp} (color online) Inclusive \j \vpt  for non-central (20\%--60\%) \pb collisions at \snn~=~2.76 TeV (see text for details on uncertainties). The used \pt ranges are: 0--2, 2--4, 4--6 and 6--10 GeV/$c$. Calculations from two transport models~\cite{Liu:2009gx} and~\cite{Zhao2013611c} are also shown (see text for details).}
\end{figure}

To allow a direct comparison with current model calculations, the inclusive \j \vpt was also calculated in a broader centrality range, namely 20\%--60\%, and it is shown in Fig.~\ref{fig:v2ptcomp}.   
In this broader centrality range, the measured \vv signal in the \pt range 2--4 GeV/$c$ deviates from zero by 2$\sigma$.
The same trend of \vpt is observed in the 20\%--60\% and in the 20\%--40\% centrality classes. 
This trend seems qualitatively different from that of the STAR measurement~\cite{Adamczyk:2012pw} at lower collision energy, which is compatible with zero for $p_{\rm T} \geq 2$ GeV/$c$ albeit in somewhat different (10\%--40\% and 0\%--80\%) centrality ranges.
Also shown in Fig.~\ref{fig:v2ptcomp} are two transport model calculations that include a J/$\psi$ regeneration component from deconfined charm quarks in the medium~\cite{Liu:2009gx,Zhao2013611c}. 
In both models about $30\%$ of the measured \j in the 20\%--60\% centrality range are regenerated. 
First, thermalized charm quarks in the medium transfer a significant elliptic flow to regenerated J/$\psi$. 
Second, primordial \j emitted out-of-plane traverse a longer path through the medium than those emitted in-plane resulting in a small apparent $v_{2}$. 
The predicted maximum \vv at $p_{\rm T}\sim2.5$~GeV/$c$ results from an interplay between the regeneration component, dominant at lower $p_{\rm T}$, and the primordial \j component which takes over at higher $p_{\rm T}$.
The first model~\cite{Liu:2009gx} is shown for the hypothesis of thermalization (full line) and non-thermalization (dashed line) of b quarks. The LHCb Collaboration measured the fraction of \j from B hadron decays in pp collisions at $\sqrt{\rm s}$ = 2.76 and 7 TeV~\cite{Aaij:2012asz,Aaij:2011jh} in the rapidity acceptance used for this measurement. 
At 7 TeV this fraction increases from 7\% at  $p_{\rm T} \sim 0$ to 15\% at $p_{\rm T} \sim 7$ GeV/$c$, while at 2.76 TeV it is about  7\% for $p_{\rm T}<12$ GeV/$c$.
In Pb-Pb collisions this fraction could increase up to $11\%$ if the B hadron \raa = 1.
If b quarks do thermalize then their elliptic flow will be transferred to B mesons at hadronization and to the \j at the B meson decay.  
In the second model~\cite{Zhao2013611c} (dash-dotted line) only the case assuming b quark thermalization is shown. 
Both models qualitatively describe the \pt dependence of the \vv and the \raa~of inclusive \j~\cite{Abelev:2012rv}. 

In summary, we reported the ALICE measurement of inclusive \j elliptic flow in the range $0 \leq p_{\rm T} < 10$ GeV/$c$ at forward rapidity in \pb collisions at \snn=2.76 TeV. 
For semi-central collisions indications of a non-zero \j \vv are observed in the intermediate \pt range. % in contrast to the zero \vv observed at RHIC energies. 
This measurement complements the results on the \j \raa, where a smaller suppression was seen at low \pt at the LHC compared to RHIC.  
Both results seem in agreement with the global picture in which a significant fraction of the observed \j is produced from deconfined charm quarks in the QGP phase.
%These results suggest that a significant fraction of the observed \j is produced from deconfined charm quarks in the QGP phase.                %%%%%%%%%%% put the body of the article here

%
%%%%%%%% acknowledgements
\newenvironment{acknowledgement}{\relax}{\relax}
\begin{acknowledgement}
\section*{Acknowledgements}
The ALICE collaboration would like to thank all its engineers and technicians for their invaluable contributions to the construction of the experiment and the CERN accelerator teams for the outstanding performance of the LHC complex.
\\
The ALICE collaboration acknowledges the following funding agencies for their support in building and
running the ALICE detector:
 \\
State Committee of Science,  World Federation of Scientists (WFS)
and Swiss Fonds Kidagan, Armenia,
 \\
Conselho Nacional de Desenvolvimento Cient\'{\i}fico e Tecnol\'{o}gico (CNPq), Financiadora de Estudos e Projetos (FINEP),
Funda\c{c}\~{a}o de Amparo \`{a} Pesquisa do Estado de S\~{a}o Paulo (FAPESP);
 \\
National Natural Science Foundation of China (NSFC), the Chinese Ministry of Education (CMOE)
and the Ministry of Science and Technology of China (MSTC);
 \\
Ministry of Education and Youth of the Czech Republic;
 \\
Danish Natural Science Research Council, the Carlsberg Foundation and the Danish National Research Foundation;
 \\
The European Research Council under the European Community's Seventh Framework Programme;
 \\
Helsinki Institute of Physics and the Academy of Finland;
 \\
French CNRS-IN2P3, the `Region Pays de Loire', `Region Alsace', `Region Auvergne' and CEA, France;
 \\
German BMBF and the Helmholtz Association;
\\
General Secretariat for Research and Technology, Ministry of
Development, Greece;
\\
Hungarian OTKA and National Office for Research and Technology (NKTH);
 \\
Department of Atomic Energy and Department of Science and Technology of the Government of India;
 \\
Istituto Nazionale di Fisica Nucleare (INFN) and Centro Fermi -
Museo Storico della Fisica e Centro Studi e Ricerche "Enrico
Fermi", Italy;
 \\
MEXT Grant-in-Aid for Specially Promoted Research, Ja\-pan;
 \\
Joint Institute for Nuclear Research, Dubna;
 \\
%Korea Foundation for International Cooperation of Science and Technology (KICOS);
National Research Foundation of Korea (NRF);
 \\
CONACYT, DGAPA, M\'{e}xico, ALFA-EC and the EPLANET Program
(European Particle Physics Latin American Network)
 \\
Stichting voor Fundamenteel Onderzoek der Materie (FOM) and the Nederlandse Organisatie voor Wetenschappelijk Onderzoek (NWO), Netherlands;
 \\
Research Council of Norway (NFR);
 \\
Polish Ministry of Science and Higher Education;
 \\
National Authority for Scientific Research - NASR (Autoritatea Na\c{t}ional\u{a} pentru Cercetare \c{S}tiin\c{t}ific\u{a} - ANCS);
 \\
Ministry of Education and Science of Russian Federation, Russian
Academy of Sciences, Russian Federal Agency of Atomic Energy,
Russian Federal Agency for Science and Innovations and The Russian
Foundation for Basic Research;
 \\
Ministry of Education of Slovakia;
 \\
Department of Science and Technology, South Africa;
 \\
CIEMAT, EELA, Ministerio de Econom\'{i}a y Competitividad (MINECO) of Spain, Xunta de Galicia (Conseller\'{\i}a de Educaci\'{o}n),
CEA\-DEN, Cubaenerg\'{\i}a, Cuba, and IAEA (International Atomic Energy Agency);
 \\
Swedish Research Council (VR) and Knut $\&$ Alice Wallenberg
Foundation (KAW);
 \\
Ukraine Ministry of Education and Science;
 \\
United Kingdom Science and Technology Facilities Council (STFC);
 \\
The United States Department of Energy, the United States National
Science Foundation, the State of Texas, and the State of Ohio.
    %%%%%%% get the lates version before submitting
\end{acknowledgement}

\bibliographystyle{apsrev4-1}
\newpage
%
%\input{}               %%%%%%%%%%% put your appendices here
%
%%%%%%%%% appendix with author list
\appendix
\section{The ALICE Collaboration}
\label{app:collab}

% Collaboration: CERN-LHC-ALICE
% Generation Date is 12 Mar 2013 00:00:22 GMT
\begingroup
\small
\begin{flushleft}
E.~Abbas\Irefn{org36632}\And
B.~Abelev\Irefn{org1234}\And
J.~Adam\Irefn{org1274}\And
D.~Adamov\'{a}\Irefn{org1283}\And
A.M.~Adare\Irefn{org1260}\And
M.M.~Aggarwal\Irefn{org1157}\And
G.~Aglieri~Rinella\Irefn{org1192}\And
M.~Agnello\Irefn{org1313}\textsuperscript{,}\Irefn{org1017688}\And
A.G.~Agocs\Irefn{org1143}\And
A.~Agostinelli\Irefn{org1132}
% ...
\renewcommand\labelenumi{\textsuperscript{\theenumi}~}
\section*{Affiliation notes}
\renewcommand\theenumi{\roman{enumi}}
\begin{Authlist}
\item \Adef{M.V.Lomonosov Moscow State University, D.V.Skobeltsyn Institute of Nuclear Physics, Moscow, Russia}Also at: M.V.Lomonosov Moscow State University, D.V.Skobeltsyn Institute of Nuclear Physics, Moscow, Russia
\item \Adef{University of Belgrade, Faculty of Physics and Vinvca Institute of Nuclear Sciences, Belgrade, Serbia}Also at: University of Belgrade, Faculty of Physics and Vinvca Institute of Nuclear Sciences, Belgrade, Serbia
\item \Adef{Institute of Theoretical Physics, University of Wroclaw, Wroclaw, Poland}Also at: Institute of Theoretical Physics, University of Wroclaw, Wroclaw, Poland
\end{Authlist}
\section*{Collaboration Institutes}
\renewcommand\theenumi{\arabic{enumi}~}
\begin{Authlist}
\item \Idef{org36632}Academy of Scientific Research and Technology (ASRT), Cairo, Egypt
\item \Idef{org1332}A. I. Alikhanyan National Science Laboratory (Yerevan Physics Institute) Foundation, Yerevan, Armenia
\item \Idef{org1279}Benem\'{e}rita Universidad Aut\'{o}noma de Puebla, Puebla, Mexico
\item \Idef{org1220}Bogolyubov Institute for Theoretical Physics, Kiev, Ukraine
\item \Idef{org20959}Bose Institute, Department of Physics and Centre for Astroparticle Physics and Space Science (CAPSS), Kolkata, India
\item \Idef{org1262}Budker Institute for Nuclear Physics, Novosibirsk, Russia
\item \Idef{org1292}California Polytechnic State University, San Luis Obispo, California, United States
\item \Idef{org1329}Central China Normal University, Wuhan, China
\item \Idef{org14939}Centre de Calcul de l'IN2P3, Villeurbanne, France
\item \Idef{org1197}Centro de Aplicaciones Tecnol\'{o}gicas y Desarrollo Nuclear (CEADEN), Havana, Cuba
\item \Idef{org1242}Centro de Investigaciones Energ\'{e}ticas Medioambientales y Tecnol\'{o}gicas (CIEMAT), Madrid, Spain
\item \Idef{org1244}Centro de Investigaci\'{o}n y de Estudios Avanzados (CINVESTAV), Mexico City and M\'{e}rida, Mexico
\item \Idef{org1335}Centro Fermi - Museo Storico della Fisica e Centro Studi e Ricerche ``Enrico Fermi'', Rome, Italy
\item \Idef{org17347}Chicago State University, Chicago, United States
\item \Idef{org1288}Commissariat \`{a} l'Energie Atomique, IRFU, Saclay, France
\item \Idef{org15782}COMSATS Institute of Information Technology (CIIT), Islamabad, Pakistan
\item \Idef{org1294}Departamento de F\'{\i}sica de Part\'{\i}culas and IGFAE, Universidad de Santiago de Compostela, Santiago de Compostela, Spain
\item \Idef{org1106}Department of Physics Aligarh Muslim University, Aligarh, India
\item \Idef{org1121}Department of Physics and Technology, University of Bergen, Bergen, Norway
\item \Idef{org1162}Department of Physics, Ohio State University, Columbus, Ohio, United States
\item \Idef{org1300}Department of Physics, Sejong University, Seoul, South Korea
\item \Idef{org1268}Department of Physics, University of Oslo, Oslo, Norway
\item \Idef{org1315}Dipartimento di Fisica dell'Universit\`{a} and Sezione INFN, Trieste, Italy
\item \Idef{org1145}Dipartimento di Fisica dell'Universit\`{a} and Sezione INFN, Cagliari, Italy
\item \Idef{org1312}Dipartimento di Fisica dell'Universit\`{a} and Sezione INFN, Turin, Italy
\item \Idef{org1285}Dipartimento di Fisica dell'Universit\`{a} `La Sapienza' and Sezione INFN, Rome, Italy
\item \Idef{org1154}Dipartimento di Fisica e Astronomia dell'Universit\`{a} and Sezione INFN, Catania, Italy
\item \Idef{org1132}Dipartimento di Fisica e Astronomia dell'Universit\`{a} and Sezione INFN, Bologna, Italy
\item \Idef{org1270}Dipartimento di Fisica e Astronomia dell'Universit\`{a} and Sezione INFN, Padova, Italy
\item \Idef{org1290}Dipartimento di Fisica `E.R.~Caianiello' dell'Universit\`{a} and Gruppo Collegato INFN, Salerno, Italy
\item \Idef{org1103}Dipartimento di Scienze e Innovazione Tecnologica dell'Universit\`{a} del Piemonte Orientale and Gruppo Collegato INFN, Alessandria, Italy
\item \Idef{org1114}Dipartimento Interateneo di Fisica `M.~Merlin' and Sezione INFN, Bari, Italy
\item \Idef{org1237}Division of Experimental High Energy Physics, University of Lund, Lund, Sweden
\item \Idef{org1192}European Organization for Nuclear Research (CERN), Geneva, Switzerland
\item \Idef{org1227}Fachhochschule K\"{o}ln, K\"{o}ln, Germany
\item \Idef{org1122}Faculty of Engineering, Bergen University College, Bergen, Norway
\item \Idef{org1136}Faculty of Mathematics, Physics and Informatics, Comenius University, Bratislava, Slovakia
\item \Idef{org1274}Faculty of Nuclear Sciences and Physical Engineering, Czech Technical University in Prague, Prague, Czech Republic
\item \Idef{org1229}Faculty of Science, P.J.~\v{S}af\'{a}rik University, Ko\v{s}ice, Slovakia
\item \Idef{org1184}Frankfurt Institute for Advanced Studies, Johann Wolfgang Goethe-Universit\"{a}t Frankfurt, Frankfurt, Germany
\item \Idef{org1215}Gangneung-Wonju National University, Gangneung, South Korea
\item \Idef{org20958}Gauhati University, Department of Physics, Guwahati, India
\item \Idef{org1212}Helsinki Institute of Physics (HIP) and University of Jyv\"{a}skyl\"{a}, Jyv\"{a}skyl\"{a}, Finland
\item \Idef{org1203}Hiroshima University, Hiroshima, Japan
\item \Idef{org1254}Indian Institute of Technology Bombay (IIT), Mumbai, India
\item \Idef{org36378}Indian Institute of Technology Indore, Indore, India (IITI)
\item \Idef{org1266}Institut de Physique Nucl\'{e}aire d'Orsay (IPNO), Universit\'{e} Paris-Sud, CNRS-IN2P3, Orsay, France
\item \Idef{org1277}Institute for High Energy Physics, Protvino, Russia
\item \Idef{org1249}Institute for Nuclear Research, Academy of Sciences, Moscow, Russia
\item \Idef{org1320}Nikhef, National Institute for Subatomic Physics and Institute for Subatomic Physics of Utrecht University, Utrecht, Netherlands
\item \Idef{org1250}Institute for Theoretical and Experimental Physics, Moscow, Russia
\item \Idef{org1230}Institute of Experimental Physics, Slovak Academy of Sciences, Ko\v{s}ice, Slovakia
\item \Idef{org1127}Institute of Physics, Bhubaneswar, India
\item \Idef{org1275}Institute of Physics, Academy of Sciences of the Czech Republic, Prague, Czech Republic
\item \Idef{org1139}Institute of Space Sciences (ISS), Bucharest, Romania
\item \Idef{org27399}Institut f\"{u}r Informatik, Johann Wolfgang Goethe-Universit\"{a}t Frankfurt, Frankfurt, Germany
\item \Idef{org1185}Institut f\"{u}r Kernphysik, Johann Wolfgang Goethe-Universit\"{a}t Frankfurt, Frankfurt, Germany
\item \Idef{org1177}Institut f\"{u}r Kernphysik, Technische Universit\"{a}t Darmstadt, Darmstadt, Germany
\item \Idef{org1256}Institut f\"{u}r Kernphysik, Westf\"{a}lische Wilhelms-Universit\"{a}t M\"{u}nster, M\"{u}nster, Germany
\item \Idef{org1246}Instituto de Ciencias Nucleares, Universidad Nacional Aut\'{o}noma de M\'{e}xico, Mexico City, Mexico
\item \Idef{org1247}Instituto de F\'{\i}sica, Universidad Nacional Aut\'{o}noma de M\'{e}xico, Mexico City, Mexico
\item \Idef{org1308}Institut Pluridisciplinaire Hubert Curien (IPHC), Universit\'{e} de Strasbourg, CNRS-IN2P3, Strasbourg, France
\item \Idef{org1182}Joint Institute for Nuclear Research (JINR), Dubna, Russia
\item \Idef{org1199}Kirchhoff-Institut f\"{u}r Physik, Ruprecht-Karls-Universit\"{a}t Heidelberg, Heidelberg, Germany
\item \Idef{org20954}Korea Institute of Science and Technology Information, Daejeon, South Korea
\item \Idef{org1017642}KTO Karatay University, Konya, Turkey
\item \Idef{org1160}Laboratoire de Physique Corpusculaire (LPC), Clermont Universit\'{e}, Universit\'{e} Blaise Pascal, CNRS--IN2P3, Clermont-Ferrand, France
\item \Idef{org1194}Laboratoire de Physique Subatomique et de Cosmologie (LPSC), Universit\'{e} Joseph Fourier, CNRS-IN2P3, Institut Polytechnique de Grenoble, Grenoble, France
\item \Idef{org1187}Laboratori Nazionali di Frascati, INFN, Frascati, Italy
\item \Idef{org1232}Laboratori Nazionali di Legnaro, INFN, Legnaro, Italy
\item \Idef{org1125}Lawrence Berkeley National Laboratory, Berkeley, California, United States
\item \Idef{org1234}Lawrence Livermore National Laboratory, Livermore, California, United States
\item \Idef{org1251}Moscow Engineering Physics Institute, Moscow, Russia
\item \Idef{org1322}National Centre for Nuclear Studies, Warsaw, Poland
\item \Idef{org1140}National Institute for Physics and Nuclear Engineering, Bucharest, Romania
\item \Idef{org1017626}National Institute of Science Education and Research, Bhubaneswar, India
\item \Idef{org1165}Niels Bohr Institute, University of Copenhagen, Copenhagen, Denmark
\item \Idef{org1109}Nikhef, National Institute for Subatomic Physics, Amsterdam, Netherlands
\item \Idef{org1283}Nuclear Physics Institute, Academy of Sciences of the Czech Republic, \v{R}e\v{z} u Prahy, Czech Republic
\item \Idef{org1264}Oak Ridge National Laboratory, Oak Ridge, Tennessee, United States
\item \Idef{org1189}Petersburg Nuclear Physics Institute, Gatchina, Russia
\item \Idef{org1170}Physics Department, Creighton University, Omaha, Nebraska, United States
\item \Idef{org1157}Physics Department, Panjab University, Chandigarh, India
\item \Idef{org1112}Physics Department, University of Athens, Athens, Greece
\item \Idef{org1152}Physics Department, University of Cape Town and  iThemba LABS, National Research Foundation, Somerset West, South Africa
\item \Idef{org1209}Physics Department, University of Jammu, Jammu, India
\item \Idef{org1207}Physics Department, University of Rajasthan, Jaipur, India
\item \Idef{org1200}Physikalisches Institut, Ruprecht-Karls-Universit\"{a}t Heidelberg, Heidelberg, Germany
\item \Idef{org1017688}Politecnico di Torino, Turin, Italy
\item \Idef{org1325}Purdue University, West Lafayette, Indiana, United States
\item \Idef{org1281}Pusan National University, Pusan, South Korea
\item \Idef{org1176}Research Division and ExtreMe Matter Institute EMMI, GSI Helmholtzzentrum f\"ur Schwerionenforschung, Darmstadt, Germany
\item \Idef{org1334}Rudjer Bo\v{s}kovi\'{c} Institute, Zagreb, Croatia
\item \Idef{org1298}Russian Federal Nuclear Center (VNIIEF), Sarov, Russia
\item \Idef{org1252}Russian Research Centre Kurchatov Institute, Moscow, Russia
\item \Idef{org1224}Saha Institute of Nuclear Physics, Kolkata, India
\item \Idef{org1130}School of Physics and Astronomy, University of Birmingham, Birmingham, United Kingdom
\item \Idef{org1338}Secci\'{o}n F\'{\i}sica, Departamento de Ciencias, Pontificia Universidad Cat\'{o}lica del Per\'{u}, Lima, Peru
\item \Idef{org1155}Sezione INFN, Catania, Italy
\item \Idef{org1313}Sezione INFN, Turin, Italy
\item \Idef{org1271}Sezione INFN, Padova, Italy
\item \Idef{org1133}Sezione INFN, Bologna, Italy
\item \Idef{org1146}Sezione INFN, Cagliari, Italy
\item \Idef{org1316}Sezione INFN, Trieste, Italy
\item \Idef{org1115}Sezione INFN, Bari, Italy
\item \Idef{org1286}Sezione INFN, Rome, Italy
\item \Idef{org36377}Nuclear Physics Group, STFC Daresbury Laboratory, Daresbury, United Kingdom
\item \Idef{org1258}SUBATECH, Ecole des Mines de Nantes, Universit\'{e} de Nantes, CNRS-IN2P3, Nantes, France
\item \Idef{org35706}Suranaree University of Technology, Nakhon Ratchasima, Thailand
\item \Idef{org1304}Technical University of Split FESB, Split, Croatia
\item \Idef{org1017659}Technische Universit\"{a}t M\"{u}nchen, Munich, Germany
\item \Idef{org1168}The Henryk Niewodniczanski Institute of Nuclear Physics, Polish Academy of Sciences, Cracow, Poland
\item \Idef{org17361}The University of Texas at Austin, Physics Department, Austin, TX, United States
\item \Idef{org1173}Universidad Aut\'{o}noma de Sinaloa, Culiac\'{a}n, Mexico
\item \Idef{org1296}Universidade de S\~{a}o Paulo (USP), S\~{a}o Paulo, Brazil
\item \Idef{org1149}Universidade Estadual de Campinas (UNICAMP), Campinas, Brazil
\item \Idef{org1239}Universit\'{e} de Lyon, Universit\'{e} Lyon 1, CNRS/IN2P3, IPN-Lyon, Villeurbanne, France
\item \Idef{org1205}University of Houston, Houston, Texas, United States
\item \Idef{org1222}University of Tennessee, Knoxville, Tennessee, United States
\item \Idef{org1310}University of Tokyo, Tokyo, Japan
\item \Idef{org1318}University of Tsukuba, Tsukuba, Japan
\item \Idef{org21360}Eberhard Karls Universit\"{a}t T\"{u}bingen, T\"{u}bingen, Germany
\item \Idef{org1225}Variable Energy Cyclotron Centre, Kolkata, India
\item \Idef{org1017687}Vestfold University College, Tonsberg, Norway
\item \Idef{org1306}V.~Fock Institute for Physics, St. Petersburg State University, St. Petersburg, Russia
\item \Idef{org1323}Warsaw University of Technology, Warsaw, Poland
\item \Idef{org1179}Wayne State University, Detroit, Michigan, United States
\item \Idef{org1143}Wigner Research Centre for Physics, Hungarian Academy of Sciences, Budapest, Hungary
\item \Idef{org1260}Yale University, New Haven, Connecticut, United States
\item \Idef{org15649}Yildiz Technical University, Istanbul, Turkey
\item \Idef{org1301}Yonsei University, Seoul, South Korea
\item \Idef{org1327}Zentrum f\"{u}r Technologietransfer und Telekommunikation (ZTT), Fachhochschule Worms, Worms, Germany
\end{Authlist}
\endgroup

  %%%%%%% get the latest version before submitting
%
%\input{}              %%%%%%%%%%%% put the references here
%
\end{document}